\documentclass[%
reprint,
amsmath,amssymb,
aps, 
superscriptaddress,
]{revtex4-2}
\usepackage{mdframed}
\usepackage{bbold}

\usepackage{amsfonts}
\usepackage{amsthm}
\usepackage{amsbsy}
\usepackage{array}
\usepackage[breaklinks=true]{hyperref}

\usepackage{lipsum}

\usepackage{mathtools}
\usepackage{romanbar}

\usepackage{dsdshorthand}
\usepackage{xcolor}

\usepackage[vcentermath]{youngtab}

\renewcommand\vol{\mathop{\mathrm{vol}}}

\newcommand\ND[1]{{\color[rgb]{0.6,0.0,0.6} [ND: #1]}}

\definecolor{ZKgreen}{rgb}{0,0.6,0.3}

\newcommand{\lla}{\llangle}
\newcommand{\rra}{\rrangle}

\newcommand{\lnsp}{\hspace{-0.5pt}}

\makeatletter
\def\@fpheader{\ }
\makeatother

\renewcommand{\bD}{{\mathbb{D}}}

\renewcommand{\cN}{{\mathcal{N}}}

\newcommand{\beq}{\begin{equation}}
\newcommand{\eeq}{\end{equation}}
\newcommand{\bal}{\begin{equation}\begin{aligned}{}}
\newcommand{\eal}{\end{aligned} \end{equation}}
\newcommand{\llangle}{\big\langle\hspace{-1.2mm}\big\langle}
\newcommand{\rrangle}{\big\rangle\hspace{-1.2mm}\big\rangle}
\DeclareMathOperator{\Li}{Li}

\begin{document}

\title{Nonlinearly Realised Defect Symmetries and Anomalies}

\author{Nadav Drukker}
\email{nadav.drukker@gmail.com}
\affiliation{Department of Mathematics, King's College London, London, WC2R 2LS, United Kingdom}
\author{Ziwen Kong}
\email{zwn.kong@gmail.com}
\affiliation{Deutsches Elektronen-Synchrotron DESY, Notkestr.\ 85, 22607 Hamburg, Germany}
\author{Petr Kravchuk}
\email{petr.kravchuk@kcl.ac.uk}
\affiliation{Department of Mathematics, King's College London, London, WC2R 2LS, United Kingdom}

\begin{abstract}
Conformal defects---extended objects in conformal field theories---carry localised excitations inherited from symmetry currents, known as the displacements and tilts. They capture the linear response of the defect to deformations of its shape or of its profile along internal symmetry directions. There is no universal formula for deformations beyond the linear order and this is subject to ambiguities of coordinate choices on coset spaces and scheme dependence in the quantum theory. We analyse the exact match between the two and identify the scheme independent quantities capturing nonlinearly realised symmetries. This leads to universal integral identities for correlation functions in the presence of tilts and displacements. We present several applications of them. We also study possible anomalies, recovering known ones, finding new expressions for them, and uncovering new ones.
\end{abstract}

\maketitle

\section{Introduction and summary}

Symmetries are of paramount importance in quantum field theory (QFT), where correlation functions satisfy Ward--Takahashi identities even for broken symmetries. In the presence of extended objects, like line or surface operators, commonly referred to as \emph{defects}, some space--time and internal symmetries are often broken. This breaking is represented to linear order by particular operators localised on the defect, known as the displacement $\bD_r$ and tilt $t_i$ operators.

Ward--Takahashi identities for symmetries broken by defects relate therefore correlation functions to ones with an integral of an extra displacement or tilt \cite{Billo:2016cpy, Padayasi:2021sik}. This is all well at the linear level in an infinitesimal deformation, but mostly unexplored beyond that. Trying to exponentiate this action leads to multiple operator insertions and requires an understanding of what happens when they collide.

We present here a formalism for dealing with multipoint correlation functions of displacement and tilt operators in conformal field theories (CFTs), summarising the main ideas and results of \cite{whale}. Our calculations allow for any renormalization scheme 
\footnote{In practice we actually restrict to symmetric schemes without unnecessary contact terms. This could be relaxed, but we see no good reason to do that.}
and carefully distinguish quantities that are scheme independent. For example, the double integral of the four-point function of tilt operators defined in \eqref{eq:tiltdefn} is sensitive to the triple operator product expansion (OPE) $t\in t\times t\times t$. At coincident points, this OPE is scheme dependent, but we identify a scheme independent part in the integral that  participates in the constraint~\eqref{eq:RresultExternal}.

When discussing symmetries, one should allow for the possibility of anomalies. While relations between displacement correlators and submanifold anomalies have been discussed in the past \cite{Bianchi:2015liz, Herzog:2017xha, Herzog:2021spv}, we initiate a systematic study and find a variety of new anomalies. As an example, for a line defect with fully broken $G$, the identity contribution $\mathbb{1}\in t\times t\times t$ is related to the topology of the loop group $LG$; see~\eqref{eq:tttanomaly}.

We stress that the relations we find are universally applicable. For a QFT without defects, the local currents have OPE representing the Lie algebra of the global symmetries and their possible anomalies. Our analysis is the analogue of that in the presence of defects; the roles of the currents is taken by tilts and displacements that are less constrained and hence the richness of the structures we uncover.

\paragraph*{Comment:}
During the course of this work, the papers \cite{Gabai:2025zcs, Belton:2025hbu, Gabai:2025hwf, Kong:2025sbk, Girault:2025kzt, 
Pozzi:2025goj,Belton:2025ief} appeared, which overlap with some of the results presented here. For earlier related work, see~\cite{Drukker:2022pxk, Gabai:2023lax}.

\subsection{Setup}
The most symmetric spherical/planar $p$-dimensional conformal defect $\cW$ in $d$-dimensions breaks the conformal group to $SO(p+1,1)\times O(d-p)$ 
\footnote{We do not include here the possibly of breaking the transverse rotation group~\cite{Gukov:2006jk, Drukker:2008jm, Kravchuk:2025evf, vortex}, though do discuss it in \cite{whale}.}.
If the theory has a global internal symmetry $G$, the defect may break it to a subgroup $H$. More generally, a defect may have arbitrary smooth geometry, so a map from $X:S^p\to S^d$ and local symmetry breaking given by $u:S^p\to G/H$.

The partition function in the presence of the defect is
\begin{align}\label{eq:Zwdefn}
    Z[X,u] \equiv \int D\f e^{-S} \cW^{X,u},
\end{align}
where the path integral is the one defining the CFT. We use $\tau^a$ for the defect coordinates and view $S^p$ as $\R^p\cup\{\infty\}$.

For the most part we are concerned with small (but finite) deformations of the symmetric defect parametrised by $v(\tau)\in\R^{d-p}$ and $w(\tau)\in\mathfrak{h}^\perp$, the part of the algebra of $G$ orthogonal to that of $H$. 
The displacement and tilt operators are defined by variational derivatives of $Z$. Specifically, their normalised connected correlators are
\begin{align}\label{eq:displacementdefn}
    &\lla \bD_{r_1}(\tau_1)\cdots \bD_{r_n}(\tau_n)\rra_c = \frac{\de}{\de v^{r_1}(\tau_1)}\cdots \frac{\de}{\de v^{r_n}(\tau_n)}\log Z\Big\vert_{v=0},
\\
\label{eq:tiltdefn}
    &\lla t_{i_1}(\tau_1)\cdots t_{i_n}(\tau_n)\rra_c = \frac{\de}{\de w^{i_1}(\tau_1)}\cdots \frac{\de}{\de w^{i_n}(\tau_n)}\log Z\Big\vert_{w=0}.
\end{align}

The partition function is invariant under global symmetry transformations, up to a possible $c$-number anomaly. For $g\in G$ and a Weyl transformation $\rho$ 
\begin{align}\label{eq:ZGanom}
    Z[ \cL_\rho(v),L_g(w)]=e^{A[\rho,v;g,w]}Z[v,w]\,.
\end{align}
Here $\cL_\rho$ and $L_g$ represent the left action of the two symmetries. Explicitly, we expand them and $A$ near the identity elements as
\begin{align}
\label{eq:elldefn}
\cL_{e^\xi}[v] &= v + \ell[\xi,v]+O(\xi^2)\,,
\\
\label{eq:Lgexpansion}
L_{e^\l}(w) &= w + l(\l,w) + O(\l^2)\,,
\\
\label{eq:Agexpansion}
A[g,w;\rho,v] &=\cA[\l,w;\xi,v]+ O(\l^2,\xi^2)\,.
\end{align}
We work to linear order in the conformal Killing vector $\xi$ and generator $\lambda\in\mathfrak{g}$ and these vector fields contain all order information in $v$ and $w$, representing in principle finite deformations of the defect.

As displacements and tilts are defined only to linear order, higher order terms in $\ell$ and $l$ are scheme-dependent. The exact map between these terms and a choice of scheme (i.e.\ contact terms in correlation functions) is non-trivial and is obtained below. We expand the vector field $\ell$ to arbitrary powers in $w(\tau)$
\begin{align}
\ell[\xi,v] &= \sum_{n=0}^\oo \frac{1}{n!}\ell_n[\xi;v,\cdots,v],
\end{align}
and likewise for $l$ and $\cA$. These coefficients satisfy relations inherited from the Lie algebras of diffeomorphisms and $G$ and $\cA$ satisfies the Wess--Zumino consistency conditions. These relations are analysed in full detail in~\cite{whale}.

Equation \eqref{eq:ZGanom}, expanded in powers of $v$ and $w$, gives relations between correlation functions of displacements and tilts and $\ell_n$, $l_n$ and $\cA_n$. In the tilt sector this is
\begin{align}\label{eq:rawtiltidsrearranged}
   \sum_{k=0}^{n}\binom{n}{k} \lla t(l_k(\l;w,\cdots, w)) t(w)^{n-k}\rra_c
   =\cA_n[\l;w,\cdots\!,w].
\end{align}
Here we introduced the notation
\begin{align}
    t(f) \equiv \int d^p\tau f^i(\tau) t_i(\tau).
\end{align}
In the following we solve for the $l_n$ in terms of the correlation functions on the left hand side, thus providing a precise mapping between geometry and scheme and furthermore, we untangle combinations of them that are scheme-independent. We present those now independently for tilts (with $v=0$) and displacements (with $w=0$). The analysis of mixed identities is in principle a straightforward extension and we leave it for future work.

\section{Tilt identities and applications}

Equations \eqref{eq:rawtiltidsrearranged} for $n=0,1$ are automatically satisfied based on the usual properties of defect CFT operators. The only subtlety is that for even $p$, the two-point function of tilts
\begin{align}
 \lla t_i(\tau_1)t_j(\tau_2)\rra
 =\frac{g_{ij}}{\tau_{12}^{2p}}
\end{align}
has an anomaly at coincident points which contributes at the next order to $\cA_2[\xi^\parallel;w,w]$---this is an anomaly under the preserved conformal transformations whose contribution to the submanifold anomaly polynomial for surface operators was discussed in~\cite{Drukker:2020dcz, Drukker:2020atp}. 

The relations at low $n$ are far less trivial when the partition function includes extra operators. In those cases it includes information about contact terms in $O\in t\times t\times O$ OPE, which leads to the integral identities of~\cite{Kutasov:1988xb, Friedan:2012hi,Drukker:2022pxk,Girault:2025kzt}, which we don't elaborate on here. Their analogues when $O=t$ are in \eqref{eq:RresultExternal}.

For $n=2$ the identity \eqref{eq:rawtiltidsrearranged} reads
\begin{align}\label{eq:ttt}
    \lla t(\l)t(w)^2\rra_c + 2\lla t(l_1(\l;w))t(w)\rra_c=\cA_2[\l;w,w].
\end{align}
Taking two variational derivatives with respect to $w$ we find
\begin{align}\label{eq:tttconstraint}
&\int d^p \tau_1 \lla t_i(\tau_1)t_j(\tau_2)t_k(\tau_3)\rra_c
+l^m_1(e_i;e_j)\lla t_m(\tau_2)t_k(\tau_3)\rra_c
\nn\\&\quad
+l^m_1(e_i;e_k)\lla t_j(\tau_2)t_m(\tau_3)\rra_c
=0+\cA_2\Big|_{\tau_2=\tau_3}\,.
\end{align}
Here $e_i$ are basis vectors of $\mathfrak{h}^\perp$. In the absence of the anomaly, the three-point function vanishes---otherwise it contributes to the beta-function of the exactly marginal tilts. This allows us to fully determine $l_1$ as
\begin{align}\label{eq:l1result}
2l_1(\l_2;\l_3)
=[\l_2,\l_3]^\perp-(\mathrm{ad}^\perp(\l_2))^*\l_3-(\mathrm{ad}^\perp(\l_3))^*\l_2,
\end{align}
where $\mathrm{ad}^\perp(\l_1)\l_2=[\l_1,\l_2]^\perp$ and $*$ denotes the adjoint with respect to the inner product~$g(\.,\.)$.

One exception are \textit{oriented} line defects, $p=1$, for which the three-point function of $t$ may be
\begin{align}\label{eq:tttseparatedlinecase}
\lla t_i(\tau_1)t_j(\tau_2)t_k(\tau_3)\rra_c=\frac{if_{ijk}}{(\tau_1-\tau_2)(\tau_2-\tau_3)(\tau_3-\tau_1)},
\end{align}
with fully antisymmetric $f_{ijk}$. Such terms are written very explicitly in a holographic setting in \cite{Bliard:2024bcz} and also naturally arise from boundary currents in WZW models.

The right-hand side of~\eqref{eq:tttseparatedlinecase} cannot be matched by the two-point functions in~\eqref{eq:tttconstraint}. Instead, it is matched by $\cA_2$ and therefore produces a non-trivial anomaly
\begin{align}
\label{eq:tttanomaly}
    \cA_2[\l;w,w]=\frac{2\pi^2i}{3}\int d\tau f_{ijk}\l^i w^j \ptl w^k,
\end{align}
with totally antisymmetric $f_{ijk}$. Indeed, an intricate cohomological analysis identifies the anomalies of $p=1$ oriented defects with $H^3(G/H,\R)$, so $H$-invariant three-forms on $\mathfrak{h}^\perp$ satisfying the cocycle condition. The cocycle condition for $f_{ijk}$ is
\begin{align}
    \label{eq:cocycle}
    [e_{[l},e_m]^kf_{ij]k}=0.
\end{align}
This can be seen as arising from Wess-Zumino consistency condition or, equivalently, a contact term in the $n=3$ version of~\eqref{eq:rawtiltidsrearranged}. Further properties of this anomaly are discussed in subsection~\ref{sec:anomalies}

At order $w^3$ \eqref{eq:rawtiltidsrearranged} reads
\begin{align}\label{eq:full_n=3_tilt}
&    \lla t(\l)t(w)^3 \rra_c+3\lla t(l_1(\l;w))t(w)^2\rra_c
    \nn\\&
    +3\lla t(l_2(\l;w,w))t(w)\rra_c=\cA_3[\l;w,w,w].
\end{align}
Functionally expanding in $w(\tau)$ and assuming $\tau_{2,3,4}$ are distinct, this is the linear Ward--Takahashi identity
\begin{align}\label{eq:ttttconstraint}
\int d^p\tau_1 \l^i \lla t_i(\tau_1) t_j(\tau_2)t_k(\tau_3)t_m(\tau_4)\rra_c=0.
\end{align}
In terms of the cross-ratios, for $p>1$, this is 
\begin{align}\label{eq:tttthomogeneous_fixed_crossratios_p>1}
    \int d^2z |z-\bar z|^{p-2}\lla t_i(0) t_j(z)t_k(1)t_m(\oo)\rra_c=0,
\end{align}
true also for any local operators replacing the unintegrated tilts.

Even this simple relation has powerful consequences. The four-point function of the displacement multiplets on 1/2 BPS lines in $\cN=2$ theories in 4d and $\cN=4$ theories in 3d were determined to first nontrivial order at strong coupling using analytic bootstrap tools in \cite{Gimenez-Grau:2019hez} and \cite{Pozzi:2024xnu} respectively. In both cases the result is a linear combination of two functions of the cross-ratio. Imposing~\eqref{eq:ttttconstraint} fixes the linear combination to the same function as in the more supersymmetric theories ($\cN=4$ SYM and ABJM respectively), leaving a single overall factor. The same conclusion was very recently reached in~\cite{Pozzi:2025goj} based on several natural assumptions and a careful OPE analysis of these models. Using \eqref{eq:ttttconstraint}, this is a one line calculation.

At this order there are more relations. Allowing two coincident tilts in the expansion of~\eqref{eq:full_n=3_tilt}, say $\tau_2=\tau_3\neq\tau_4$ picks up contributions from the third term on the left hand side. Integrating over $\tau_1$, $\tau_2$ with an appropriate measure allows us to determine $l_2$ and find the scheme independent components in it, dictated by the choice of contact terms in $\lla tttt\rra$. We do not include the details here, just that the scheme independent part is equivalent to the identity (for $p=1$)
\begin{align}\label{eq:RresultExternal}
    R_{ijkm} =
    2\int d\tau \log |\tau| 
    \llangle t_{i} (1) t_{j} (\tau) t_k (0) t_m (\infty) \rrangle_c,  
\end{align}
where $R$ is the Riemann tensor of the defect conformal manifold. This has previously been derived in~\cite{Kutasov:1988xb, Friedan:2012hi, Drukker:2022pxk, Girault:2025kzt}, including higher $p$.

Equations \eqref{eq:tttthomogeneous_fixed_crossratios_p>1}, \eqref{eq:RresultExternal} and all integral identities below are correct with a proper distributional definition of the correlator, allowing for contact terms in the $t\times t$ OPE. The identity and $t$ contributions are behind the anomalies $\cA_2[\xi^\|;w,w]$ and \eqref{eq:tttanomaly}.

In perturbative settings there are often other operators of near integer dimension, that can appear in the OPE, say a singlet $\phi$ of dimension $p+\gamma_\phi$, such that $\delta(\tau_{12})\phi(\tau_1)\in t(\tau_1)\times t(\tau_2)$. Without accounting for these contact terms, the integral identities like \eqref{eq:tttthomogeneous_fixed_crossratios_p>1} seem to be violated \cite{Cavaglia:2022qpg}. A careful treatment does not only validate the equation, it allows to determine some relations between anomalous dimensions and structure constants \cite{Sakkas:2024dvm, Belton:2025hbu, Girault:2025kzt}. For example, the connected six-point function of tilts on the magnetic line defect in the $O(N)$ model in $d=4-\varepsilon$ \cite{Wilson:1971dc, Allais:2014fqa, Cuomo:2021kfm, Gimenez-Grau:2022czc} up to $O(\varepsilon)$ comprises only contact terms. Without calculating the smooth $O(\varepsilon^2)$ contribution, a delta function term in the six-point analogue of \eqref{eq:full_n=3_tilt} (i.e.\ the $n=5$ version of~\eqref{eq:rawtiltidsrearranged}) relates (to order $\varepsilon$)
\begin{align}
\label{eq:tttt-ttpp}
&C_t^{-1}\lla t_i(0)t_j(z)t_k(1)t_m(\infty)\rra\!
=\!\delta_{ij}\lla \phi(0)\phi(z)t_k(1)t_m(\infty)\rra
\nn\\
&\!+\!\delta_{ik}\lla \phi(0)t_j(z)\phi(1)t_m(\infty)\rra
\!+\!\delta_{im}\lla \phi(0)t_j(z)t_k(1)\phi(\infty)\rra,
\end{align}
where $g_{ij}=C_t\de_{ij}$.
Furthermore, the contact terms in this relation then imply that the structure constants satisfy $C_{\phi\phi\phi}=3C_{tt\phi}/C_t$ to order $\varepsilon$.
This matches perturbative calculations \cite{Gimenez-Grau:2022czc}, but is actually valid in any theory with a similar spectrum and perturbative structure, including (with appropriate adaptation for planarity issues) in $\cN=4$ SYM \cite{Artico:2024wut}.

Finally, if $\tau_2=\tau_3=\tau_4$ we find a contribution from the anomaly. Now we find nontrivial cohomology for $p=2$ with the anomaly given by $H^4(G/H,\R)$, i.e.\ $H$-invariant four-forms on $\mathfrak{h}^\perp$ satisfying the cocycle condition. The anomaly can be extracted from the four-point function via
\begin{align}
\label{eq:ttttanomaly}
  \O_{ijkl} = \frac{\pi^2}{4i}\int d^2 z (z-\bar z)D_{1111}
  \lla t_{[i}(\oo)t_j(z)t_k(1)t_{l]}(0)\rra_c.
\end{align}
where
\begin{align}
\label{eq:D1111}
  D_{1111}(z,\bar z)&=\frac{1}{z-\bar z}\big(2\Li_2(z)-2\Li_2(\bar z)
  \nn\\&\quad
  +\log(z\bar z)(\log(1-z)-\log(1-\bar z))\big)\,,
\end{align}
is the usual massless scalar box integral \cite{Usyukina:1992jd}. The cocycle condition for $\O$ is
\begin{align}
  [e_{[m}, e_n]^l\O_{ijk]l}=0.
\end{align}

\subsection{Anomalies}
\label{sec:anomalies}

The anomaly coefficients $f_{ijk}$ and $\O_{ijkl}$ are further constrained by global consistency and by the bulk theory. 

Firstly, we find that the leading-order anomaly terms $\cA_2$ ($p=1$) and $\cA_3$ ($p=2$) that are determined by these coefficients can be completed to all-orders (in $w$) solutions of the Wess-Zumino consistency conditions, but these solutions are necessarily singular for some values of $w$. These singularities can apparently be resolved by interpreting the partition function as a section of a non-trivial line bundle over the space of maps $\mathrm{Map}(S^p,G/H)$. This implies quantization conditions on $f_{ijk}$ and $\O_{ijkl}$.

Secondly, the non-triviality of this line bundle implies that these anomalies cannot on their own be Wess-Zumino consistent when extended to position-depednent $G$-gauge transformations. Since we generally expect that it should be possible to couple the bulk CFT to a background gauge field for $G$ even in the presence of our conformal defect, this implies that the above anomalies can be realized only in bulk CFTs with non-trivial 't Hooft anomalies for $G$. The unbroken group $H$ is restricted, by a converse argument, to sufficiently anomaly-free subgroups of $G$. See also \cite{Jensen:2017eof, Choi:2025ebk, Wen:2025xka, Copetti:2025sym}.

An example of a theory realizing a non-trivial $f_{ijk}$ is the probe string in the $AdS_3$ model in~\cite{Bliard:2024bcz}, with integrality conditions coming from properly quantized fluxes. A non-trivial $\Omega_{ijkl}$ appears in the case of the surface operator in the $\cN=(2,0)$ theory with $A_{N-1}$ algebra. There the antisymmetric component of $\lla tttt\rra$ at large $N$ is proportional to the function $D_{3333}(z,\bar z)$ times $N$ \cite{Drukker:2020swu} and the integral satisfies an integrality condition.

\section{Displacement identities and applications}

The identities in the displacement sector are formally the same as in the tilt case and read 
\begin{align}\label{eq:rawD}
    \sum_{k=0}^{n}\binom{n}{k} \lla \bD(\ell_k[\xi;v,\cdots, v]) \bD(v)^{n-k}\rra_c
    =\cA_n[\xi;v,\cdots ,v].
 \end{align}
Their analysis is largely analogous to the tilt case, with the following key differences. Firstly, since the coupling $v$ of $\bD$ has scaling dimension $-1$, its powers have to be accompanied by spacetime derivatives in power series expansions in $v$. This means that $\ell_k[\xi;v,\cdots, v]$ are not pointwise functions of their arguments, but rather local functionals of degree $k$ in derivatives. Our use of the square brackets is to highlight this distinction.

Secondly, the anomalies $\cA_n[\xi;v,\cdots,v]$ can be non-trivial for both broken and unbroken $\xi$. The unbroken component $\cA^\|$ is expressed via~\eqref{eq:rawD} in terms the failure of $\lla \bD\cdots\bD\rra_c$ to be conformally-invariant at coincident points. The broken component $\cA^\perp$ is largely analogous to the tilt case. Both the displacement anomalies $\cA^\perp$ and $\cA^\|$ descend from the Weyl and the gravitational anomalies that show up in the derivation of~\eqref{eq:ZGanom} from Weyl and diffeomorphism invariance of $Z$.

The analysis of $n=0,1,2$ versions of~\eqref{eq:rawD} is relatively uneventful. For $p$ even it reproduces via $\cA_2^\|$ the known result~\cite{Bianchi:2015liz} that the two-point function coefficient $C_\bD$, defined as
\begin{align}
    \lla \bD_r(\tau_1) \bD_s(\tau_2)\rra_c
    =\frac{C_\bD\de_{rs}}{\tau_{12}^{2(p+1)}}\,,
\end{align}
is related to a Weyl anomaly coefficient. For odd $p$ and $d=p+1$, it reproduces via $\cA^\perp_2$ the result~\cite{Herzog:2017xha} that relates the three-point function coefficient $C_{\bD\bD\bD}$ to a Weyl anomaly coefficient. The only novelty here is the systematic nature of our approach and that we can obtain formulas valid for any $p$.

An additional result worth mentioning is related to an exotic tensor structure in $\lla \bD\bD\bD\rra$, similar to~\eqref{eq:tttseparatedlinecase}, that is allowed for $p=1$ and $d=4$ and is proportional to the 3d Levi--Civita symbol. We find that the $n=2$ identity sets its coefficient to zero.
    
The $n=3$ displacement identity reads
\begin{align}\label{eq:DDDDidentity}
    \lla \bD(\xi^\perp) \bD(v)^{3}\rra_c
    +3\lla \bD(\ell_2[\xi;v,v]) \bD(v)\rra_c
    =\cA_3[\xi;v,v,v].
\end{align}
Functionally expanding in $v^r(\tau)$ for distinct $\tau_{2,3,4}$ gives an equation similar to \eqref{eq:ttttconstraint} with three measures (and $\hat e_1$ a unit $p$-vector)
\begin{align}
\label{eq:DDDDnolog}
\int d^p\tau f_n(\tau_1)\lla \bD_r(\tau) \bD_s(\hat e_1)\bD_t(0)\bD_u(\infty)\rra_c=0\,,
\\
f_0(\tau)=1,
\quad
f_1(\tau)=\tau\.\hat e_1,
\quad
f_2(\tau) = \tau^2.
\end{align}
The different $f_0,f_1,f_2$ come from the different choices of the conformal Killing vector $\xi$. For $p=2$ an additional measure $f_-(\tau) = \tau\.\hat e_2$ is available, where $\hat e_2$ is a unit vector orthogonal to $\hat e_1$.

Equation \eqref{eq:DDDDidentity} at two coincident points captures contact terms contributions to $\bD\in\bD\times\bD\times\bD$. 
The scheme-independent content is captured by the tensors
\begin{widetext}
\begin{align}\label{eq:Rnresult}
    R^n_{r\lnsp stu}&\equiv \half \vol S^{p-1}\int d^p\tau \log|\tau| f_n(\tau)
    \lla \bD_r(\tau)\bD_s(\hat e_1)\bD_t(0)\bD_u(\oo)\rra_c,\\
\label{eq:Rminus}
R^-_{r\lnsp stu}&\equiv -\frac{i\pi}{6}\int d^2\tau \log\left|\frac{\tau}{(\tau-1)^2}\right| (\tau-\bar\tau)
    \lla \bD_r(\tau)\bD_s(\hat e_1)\bD_t(0)\bD_u(\oo)\rra_c,
\end{align}
\end{widetext}
where $R^-$ is only defined for $p=2$.

We find that~\eqref{eq:DDDDidentity} fully constrains $R^n$ to be
{\allowdisplaybreaks
\begin{align}
\label{eq:RnfromIdentity}
R^0_{r\lnsp stu} &= -(p+1)C_\bD \de_{r\lnsp s}\de_{tu}+C_\bD (\de_{rt}\de_{su}-\de_{ru}\de_{st}),
\\
R^1_{r\lnsp stu} &= -\frac{p}{2}C_\bD (\de_{rt}\de_{su}-\de_{ru}\de_{st}),
\\
R^2_{r\lnsp stu} &= (p+1)C_\bD \de_{r\lnsp s}\de_{tu}+C_\bD (\de_{rt}\de_{su}-\de_{ru}\de_{st}),
\label{eq:RnfromIdentity_last}
\end{align}}
and $R^-=0$. These are the analogues of~\eqref{eq:RresultExternal}. Due to the symmetry under transverse rotations, as well as the permutation symmetries of $R^n$ and $R^-$ which are implicit in the definition~\eqref{eq:Rnresult}, the identities for $R^n$ imply 3 scalar equations which are all contained in $R^0$. For $p=d-4$ there is an extra equation expressing the vanishing of $\e_{r\lnsp stu}$ structure in $R^0$. For $p=2$, $R^-$ gives another scalar equation.

If one generalizes~\eqref{eq:RnfromIdentity}-\eqref{eq:RnfromIdentity_last} to the case of broken transverse rotations, the $R^n$ identities become equivalent to four rank-4 tensor identities, where one tensor is totally symmetric, one is totally anti-symmetric, and two have the symmetries of the Riemann tensor. For $p=2$, $R^-$ gives an additional identity with symmetries of the Riemann tensor.

Correlators of displacements were computed in a variety of models~\cite{Giombi:2017cqn, Giombi:2018qox, Liendo:2018ukf, Kiryu:2018phb, Beccaria:2019dws, Gimenez-Grau:2019hez, Bianchi:2020hsz, Drukker:2020swu, Ferrero:2021bsb, Giombi:2021zfb, Giombi:2022anm, Cavaglia:2022qpg, Gimenez-Grau:2022czc, Bliard:2023zpe, Ferrero:2023znz, Ferrero:2023gnu, Artico:2024wut, Pozzi:2024xnu, Bliard:2024bcz} and they all satisfy the identities \footnote{As already noted in \cite{Gabai:2025zcs, Girault:2025kzt}.}. Unfortunately, we are not familiar with examples in $p=2$ where $\lla \bD\bD\bD\bD\rra$ is known and defect parity is broken so that $R^-=0$ is a nontrivial identity.

By looking at perturbative theories with multiple fields of nearly integer dimension less than $2p$, we find relations for the CFT data. For example, for the non-BPS Wilson loop of $\cN=4$ theory at weak coupling \cite{Alday:2007he, Polchinski:2011im}, with some assumptions about other fields, the anomalous dimension and structure constant for the transverse field-strength field $F_{rs}$ and for a symmetric traceless $\cD_{rs}\sim D_{(r} F_{s)\tau}$ are
\begin{align}
\gamma_F&=\frac{C_\bD}{4}+O(\lambda^2)\,,
&\quad
C_{\bD\bD F}^2&=\frac{C_\bD^3}{64}+O(\lambda^4)\,.
\\
\gamma_\cD&=\frac{5C_\bD}{12}+O(\lambda^2)\,,
&\quad
C_{\bD\bD\cD}^2&=\frac{25C_\bD^3}{192}+O(\lambda^4)\,.
\end{align}

Finally,~\eqref{eq:DDDDidentity} at three coincident points involves the identity contribution to the contact term of four displacement operators and lead to two possible anomaly terms for $p=2$ (and no anomaly terms for $p=1$)
\begin{equation}
\label{apm}
a_\pm = \int d^2 z \mu_-^{r\lnsp stu}(z,\bar z) \lla \bD_r(\oo)\bD_s(z)\bD_t(1)\bD_u(0)\rra_c,
\end{equation}
where the measures are very similar to the case of tilts~\eqref{eq:ttttanomaly},
\begin{widetext}
\begin{align}
\label{eq:mu+}
  \mu^{r\lnsp stu}_+(z,\bar z) &= \frac{\pi^3(z-\bar z)^2}{48(d-2)d}\bar{D}_{1111}(z,\bar z)\p{\de^{r\lnsp s}\de^{tu}+\de^{rt}\de^{su}+\de^{ru}\de^{st}},
\\
\label{eq:mu-}
  \mu^{r\lnsp stu}_-(z,\bar z) &= \frac{\pi^2(z-\bar z)}{48(d-2)(d-3)}\bar{D}_{1111}(z,\bar z)\p{
    (1-z-\bar z)\de^{r\lnsp s}\de^{tu}
    +
    (z+\bar z-z\bar z)\de^{rt}\de^{su}
    +
    (z\bar z-1)\de^{ru}\de^{st}
  }.
\end{align}
\end{widetext}

The coefficient $a_+$ is a linear combination of known Weyl anomaly coefficient for surface operators \cite{Deser:1993yx, Graham:1999pm}, where in the notations of \cite{Chalabi:2021jud}
\begin{align}
a_+=a_\cD^{(2d)}-\frac{2d_1^{(2d)}}{3}\,,
\end{align}

We expect that $a_-$ is related to the gravitational anomaly of $p=2$ defects. We have not found necessary results on this anomaly in the literature, and therefore are unable to fully identify $a_-$ at present. We have not found a model where $\lla \bD\bD\bD\bD\rra$ is known and $a_-$ is not trivially zero due to $z-\bar z$ factor in the measure. Note that for $d=3$ there is no non-trivial $a_-$ term in $\cA_3$.

Equation \eqref{apm} is the first expression for the $a_\cD^{(2d)}$ anomaly coefficient in terms of correlation functions of local operators. Needless to say, using the four-point function for the surface operator in the $\cN=(2,0)$ \cite{Giombi:2023vzu}, our formula gives the correct expression for the anomaly \cite{Rodgers:2018mvq, Estes:2018tnu, Jensen:2018rxu, Chalabi:2020iie, Wang:2020xkc}. We also calculated it for a surface operator in a perturbative theory with a single defect operator $s$ of dimension close to two and can express it in terms of structure constants and anomalous dimensions as
\begin{align}
a_{\cD}^{(2d)}
=\frac{\pi^2}{2}C_\bD
-\frac{\pi^4}{2}\frac{1+\gamma_s}{\gamma_s^2}C_{\bD \bD s}^2
=-\frac{\pi^2}{2}C_\bD\frac{\gamma_s}{2+\gamma_s}
+\cdots
\end{align}
This expression is correct to the order where the connected four-point function gets only contact term contributions. This expression is robust to allowing other operators in the spectrum and matches the expression for the anomaly of the $O(N)$ invariant surface operator in the $O(N)$ model (with the structure constant determined from the other identities) \cite{Trepanier:2023tvb,Giombi:2023dqs, Raviv-Moshe:2023yvq, Diatlyk:2024ngd}.

\section{Conclusions}

We presented here a collection of relations for correlation functions of tilt and displacement operators. \cite{whale} will provide a full derivation and details with fully rigorous analysis of all contact terms, which parts are scheme independent, and a precise match between a scheme choice and a choice of coordinates on the $G/H$ coset beyond the linear order.

\cite{whale} also contains a very detailed exploration of several defects, including lines and surfaces in the Wilson--Fischer $O(N)$ model, the 1/2 BPS Wilson loop of $\cN=4$ SYM in 4d, ABJM, line defects in models dual to $AdS_3\times S^3\times S^3\times S^1$ and surface operators in the $\cN=(2,0)$ theory in 6d.

Our methodology reveals constraints missed by simple application of one or two global charges, including $R^-$ \eqref{eq:Rminus} and the anomalies for tilts \eqref{eq:tttanomaly}, \eqref{eq:ttttanomaly} and displacement \eqref{apm}. It also easily extends to any $n$-point function and indeed \eqref{eq:tttt-ttpp} is derived from a six-point function identity.

The integral identities are naturally classified by the number of coincident tilts/displacements which matches the degree of transcendentality of the integration measure: flat/polynomial (\ref{eq:tttanomaly}, \ref{eq:full_n=3_tilt}, \ref{eq:ttttconstraint}, \ref{eq:DDDDnolog}), logarithmic (\ref{eq:RresultExternal}, \ref{eq:Rnresult}, \ref{eq:Rminus}) and dilogarithm (\ref{eq:D1111}, \ref{eq:ttttanomaly}, \ref{apm}). The repeated appearance of the exact same $D_{1111}$ function in the anomalies \eqref{eq:ttttanomaly} and \eqref{apm} deserves a deeper exploration.

\begin{acknowledgments}
We are grateful to L.~Bianchi, G.~Bliard, C.~Copetti, S.~Komatsu, I.~Landea, V.~Niarchos, K.~Papadodimas, M.~Preti, A.~Stergiou and G.~Watts for useful discussions and to J.~Belton for related collaboration. 
We thank J.~Barrat, P.~Ferrero and I.~Landea for Mathematica codes.
ND would like to acknowledge the hospitality of the CERN 
theory group, DESY theory group, EPFL Lausanne, 
KITP Santa Barbara, Perimeter Institute, Newton Institute and the 
Simons Center in the course of this work.
ND's research is supported in part by the Science 
Technology \& Facilities council under the
grants ST/P000258/1 and ST/X000753/1 and 
by grant NSF PHY-2309135 to the Kavli Institute for Theoretical Physics (KITP). 
ZK is supported by ERC-2021-CoG---BrokenSymmetries 101044226. 
The work of PK was funded by UK Research and Innovation (UKRI) under the UK
government's Horizon Europe funding Guarantee [grant number EP/X042618/1] and the
Science and Technology Facilities Council [grant number ST/X000753/1].\end{acknowledgments}

\bibliographystyle{apsrev4-1}
\bibliography{refs}

\begin{thebibliography}{69}%
\makeatletter
\providecommand \@ifxundefined [1]{%
 \@ifx{#1\undefined}
}%
\providecommand \@ifnum [1]{%
 \ifnum #1\expandafter \@firstoftwo
 \else \expandafter \@secondoftwo
 \fi
}%
\providecommand \@ifx [1]{%
 \ifx #1\expandafter \@firstoftwo
 \else \expandafter \@secondoftwo
 \fi
}%
\providecommand \natexlab [1]{#1}%
\providecommand \enquote  [1]{``#1''}%
\providecommand \bibnamefont  [1]{#1}%
\providecommand \bibfnamefont [1]{#1}%
\providecommand \citenamefont [1]{#1}%
\providecommand \href@noop [0]{\@secondoftwo}%
\providecommand \href [0]{\begingroup \@sanitize@url \@href}%
\providecommand \@href[1]{\@@startlink{#1}\@@href}%
\providecommand \@@href[1]{\endgroup#1\@@endlink}%
\providecommand \@sanitize@url [0]{\catcode `\\12\catcode `\$12\catcode `\&12\catcode `\#12\catcode `\^12\catcode `\_12\catcode `\%12\relax}%
\providecommand \@@startlink[1]{}%
\providecommand \@@endlink[0]{}%
\providecommand \url  [0]{\begingroup\@sanitize@url \@url }%
\providecommand \@url [1]{\endgroup\@href {#1}{\urlprefix }}%
\providecommand \urlprefix  [0]{URL }%
\providecommand \Eprint [0]{\href }%
\providecommand \doibase [0]{http://dx.doi.org/}%
\providecommand \selectlanguage [0]{\@gobble}%
\providecommand \bibinfo  [0]{\@secondoftwo}%
\providecommand \bibfield  [0]{\@secondoftwo}%
\providecommand \translation [1]{[#1]}%
\providecommand \BibitemOpen [0]{}%
\providecommand \bibitemStop [0]{}%
\providecommand \bibitemNoStop [0]{.\EOS\space}%
\providecommand \EOS [0]{\spacefactor3000\relax}%
\providecommand \BibitemShut  [1]{\csname bibitem#1\endcsname}%
\let\auto@bib@innerbib\@empty
\bibitem [{\citenamefont {Bill\`o}\ \emph {et~al.}(2016)\citenamefont {Bill\`o}, \citenamefont {Gon\c{c}alves}, \citenamefont {Lauria},\ and\ \citenamefont {Meineri}}]{Billo:2016cpy}%
  \BibitemOpen
  \bibfield  {author} {\bibinfo {author} {\bibfnamefont {M.}~\bibnamefont {Bill\`o}}, \bibinfo {author} {\bibfnamefont {V.}~\bibnamefont {Gon\c{c}alves}}, \bibinfo {author} {\bibfnamefont {E.}~\bibnamefont {Lauria}}, \ and\ \bibinfo {author} {\bibfnamefont {M.}~\bibnamefont {Meineri}},\ }\href {\doibase 10.1007/JHEP04(2016)091} {\bibfield  {journal} {\bibinfo  {journal} {JHEP}\ }\textbf {\bibinfo {volume} {04}},\ \bibinfo {pages} {091} (\bibinfo {year} {2016})},\ \Eprint {http://arxiv.org/abs/1601.02883} {arXiv:1601.02883} \BibitemShut {NoStop}%
\bibitem [{\citenamefont {Padayasi}\ \emph {et~al.}(2022)\citenamefont {Padayasi}, \citenamefont {Krishnan}, \citenamefont {Metlitski}, \citenamefont {Gruzberg},\ and\ \citenamefont {Meineri}}]{Padayasi:2021sik}%
  \BibitemOpen
  \bibfield  {author} {\bibinfo {author} {\bibfnamefont {J.}~\bibnamefont {Padayasi}}, \bibinfo {author} {\bibfnamefont {A.}~\bibnamefont {Krishnan}}, \bibinfo {author} {\bibfnamefont {M.~A.}\ \bibnamefont {Metlitski}}, \bibinfo {author} {\bibfnamefont {I.~A.}\ \bibnamefont {Gruzberg}}, \ and\ \bibinfo {author} {\bibfnamefont {M.}~\bibnamefont {Meineri}},\ }\href {\doibase 10.21468/SciPostPhys.12.6.190} {\bibfield  {journal} {\bibinfo  {journal} {SciPost Phys.}\ }\textbf {\bibinfo {volume} {12}},\ \bibinfo {pages} {190} (\bibinfo {year} {2022})},\ \Eprint {http://arxiv.org/abs/2111.03071} {arXiv:2111.03071} \BibitemShut {NoStop}%
\bibitem [{\citenamefont {Drukker}\ \emph {et~al.}(2025{\natexlab{a}})\citenamefont {Drukker}, \citenamefont {Kong},\ and\ \citenamefont {Kravchuk}}]{whale}%
  \BibitemOpen
  \bibfield  {author} {\bibinfo {author} {\bibfnamefont {N.}~\bibnamefont {Drukker}}, \bibinfo {author} {\bibfnamefont {Z.}~\bibnamefont {Kong}}, \ and\ \bibinfo {author} {\bibfnamefont {P.}~\bibnamefont {Kravchuk}},\ }\href@noop {} {\  (\bibinfo {year} {2025}{\natexlab{a}})},\ \bibinfo {note} {to appear}\BibitemShut {NoStop}%
\bibitem [{Note1()}]{Note1}%
  \BibitemOpen
  \bibinfo {note} {In practice we actually restrict to symmetric schemes without unnecessary contact terms. This could be relaxed, but we see no good reason to do that.}\BibitemShut {Stop}%
\bibitem [{\citenamefont {Bianchi}\ \emph {et~al.}(2016)\citenamefont {Bianchi}, \citenamefont {Meineri}, \citenamefont {Myers},\ and\ \citenamefont {Smolkin}}]{Bianchi:2015liz}%
  \BibitemOpen
  \bibfield  {author} {\bibinfo {author} {\bibfnamefont {L.}~\bibnamefont {Bianchi}}, \bibinfo {author} {\bibfnamefont {M.}~\bibnamefont {Meineri}}, \bibinfo {author} {\bibfnamefont {R.~C.}\ \bibnamefont {Myers}}, \ and\ \bibinfo {author} {\bibfnamefont {M.}~\bibnamefont {Smolkin}},\ }\href {\doibase 10.1007/JHEP07(2016)076} {\bibfield  {journal} {\bibinfo  {journal} {JHEP}\ }\textbf {\bibinfo {volume} {07}},\ \bibinfo {pages} {076} (\bibinfo {year} {2016})},\ \Eprint {http://arxiv.org/abs/1511.06713} {arXiv:1511.06713} \BibitemShut {NoStop}%
\bibitem [{\citenamefont {Herzog}\ and\ \citenamefont {Huang}(2017)}]{Herzog:2017xha}%
  \BibitemOpen
  \bibfield  {author} {\bibinfo {author} {\bibfnamefont {C.~P.}\ \bibnamefont {Herzog}}\ and\ \bibinfo {author} {\bibfnamefont {K.-W.}\ \bibnamefont {Huang}},\ }\href {\doibase 10.1007/JHEP10(2017)189} {\bibfield  {journal} {\bibinfo  {journal} {JHEP}\ }\textbf {\bibinfo {volume} {10}},\ \bibinfo {pages} {189} (\bibinfo {year} {2017})},\ \Eprint {http://arxiv.org/abs/1707.06224} {arXiv:1707.06224} \BibitemShut {NoStop}%
\bibitem [{\citenamefont {Herzog}\ and\ \citenamefont {Schaub}(2022)}]{Herzog:2021spv}%
  \BibitemOpen
  \bibfield  {author} {\bibinfo {author} {\bibfnamefont {C.~P.}\ \bibnamefont {Herzog}}\ and\ \bibinfo {author} {\bibfnamefont {V.}~\bibnamefont {Schaub}},\ }\href {\doibase 10.1007/JHEP01(2022)121} {\bibfield  {journal} {\bibinfo  {journal} {JHEP}\ }\textbf {\bibinfo {volume} {01}},\ \bibinfo {pages} {121} (\bibinfo {year} {2022})},\ \Eprint {http://arxiv.org/abs/2107.11604} {arXiv:2107.11604} \BibitemShut {NoStop}%
\bibitem [{\citenamefont {Gabai}\ \emph {et~al.}(2025{\natexlab{a}})\citenamefont {Gabai}, \citenamefont {Sever},\ and\ \citenamefont {Zhong}}]{Gabai:2025zcs}%
  \BibitemOpen
  \bibfield  {author} {\bibinfo {author} {\bibfnamefont {B.}~\bibnamefont {Gabai}}, \bibinfo {author} {\bibfnamefont {A.}~\bibnamefont {Sever}}, \ and\ \bibinfo {author} {\bibfnamefont {D.-l.}\ \bibnamefont {Zhong}},\ }\href {\doibase 10.1103/gsfg-wrps} {\bibfield  {journal} {\bibinfo  {journal} {Phys. Rev. D}\ }\textbf {\bibinfo {volume} {112}},\ \bibinfo {pages} {065004} (\bibinfo {year} {2025}{\natexlab{a}})},\ \Eprint {http://arxiv.org/abs/2501.06900} {arXiv:2501.06900} \BibitemShut {NoStop}%
\bibitem [{\citenamefont {Belton}\ \emph {et~al.}(2025)\citenamefont {Belton}, \citenamefont {Drukker}, \citenamefont {Kong},\ and\ \citenamefont {Stergiou}}]{Belton:2025hbu}%
  \BibitemOpen
  \bibfield  {author} {\bibinfo {author} {\bibfnamefont {J.}~\bibnamefont {Belton}}, \bibinfo {author} {\bibfnamefont {N.}~\bibnamefont {Drukker}}, \bibinfo {author} {\bibfnamefont {Z.}~\bibnamefont {Kong}}, \ and\ \bibinfo {author} {\bibfnamefont {A.}~\bibnamefont {Stergiou}},\ }\href {\doibase 10.1088/1751-8121/adf925} {\bibfield  {journal} {\bibinfo  {journal} {J. Phys. A}\ }\textbf {\bibinfo {volume} {58}},\ \bibinfo {pages} {345401} (\bibinfo {year} {2025})},\ \Eprint {http://arxiv.org/abs/2503.07710} {arXiv:2503.07710} \BibitemShut {NoStop}%
\bibitem [{\citenamefont {Gabai}\ \emph {et~al.}(2025{\natexlab{b}})\citenamefont {Gabai}, \citenamefont {Gorbenko},\ and\ \citenamefont {Qiao}}]{Gabai:2025hwf}%
  \BibitemOpen
  \bibfield  {author} {\bibinfo {author} {\bibfnamefont {B.}~\bibnamefont {Gabai}}, \bibinfo {author} {\bibfnamefont {V.}~\bibnamefont {Gorbenko}}, \ and\ \bibinfo {author} {\bibfnamefont {J.}~\bibnamefont {Qiao}},\ }\href@noop {} {\  (\bibinfo {year} {2025}{\natexlab{b}})},\ \Eprint {http://arxiv.org/abs/2508.08250} {arXiv:2508.08250} \BibitemShut {NoStop}%
\bibitem [{\citenamefont {Kong}(2025)}]{Kong:2025sbk}%
  \BibitemOpen
  \bibfield  {author} {\bibinfo {author} {\bibfnamefont {Z.}~\bibnamefont {Kong}},\ }in\ \href@noop {} {\emph {\bibinfo {booktitle} {{16th International Workshop on Lie Theory and Its Applications in Physics}}}}\ (\bibinfo {year} {2025})\ \Eprint {http://arxiv.org/abs/2509.23797} {arXiv:2509.23797} \BibitemShut {NoStop}%
\bibitem [{\citenamefont {Girault}\ \emph {et~al.}(2025)\citenamefont {Girault}, \citenamefont {Paulos},\ and\ \citenamefont {van Vliet}}]{Girault:2025kzt}%
  \BibitemOpen
  \bibfield  {author} {\bibinfo {author} {\bibfnamefont {B.}~\bibnamefont {Girault}}, \bibinfo {author} {\bibfnamefont {M.~F.}\ \bibnamefont {Paulos}}, \ and\ \bibinfo {author} {\bibfnamefont {P.}~\bibnamefont {van Vliet}},\ }\href@noop {} {\  (\bibinfo {year} {2025})},\ \Eprint {http://arxiv.org/abs/2509.26561} {arXiv:2509.26561} \BibitemShut {NoStop}%
\bibitem [{\citenamefont {Pozzi}(2025)}]{Pozzi:2025goj}%
  \BibitemOpen
  \bibfield  {author} {\bibinfo {author} {\bibfnamefont {R.~G.}\ \bibnamefont {Pozzi}},\ }\href@noop {} {\  (\bibinfo {year} {2025})},\ \Eprint {http://arxiv.org/abs/2510.06950} {arXiv:2510.06950} \BibitemShut {NoStop}%
\bibitem [{\citenamefont {Belton}\ and\ \citenamefont {Kong}(2025)}]{Belton:2025ief}%
  \BibitemOpen
  \bibfield  {author} {\bibinfo {author} {\bibfnamefont {J.}~\bibnamefont {Belton}}\ and\ \bibinfo {author} {\bibfnamefont {Z.}~\bibnamefont {Kong}},\ }\href@noop {} {\  (\bibinfo {year} {2025})},\ \Eprint {http://arxiv.org/abs/2510.08519} {arXiv:2510.08519} \BibitemShut {NoStop}%
\bibitem [{\citenamefont {Drukker}\ \emph {et~al.}(2022)\citenamefont {Drukker}, \citenamefont {Kong},\ and\ \citenamefont {Sakkas}}]{Drukker:2022pxk}%
  \BibitemOpen
  \bibfield  {author} {\bibinfo {author} {\bibfnamefont {N.}~\bibnamefont {Drukker}}, \bibinfo {author} {\bibfnamefont {Z.}~\bibnamefont {Kong}}, \ and\ \bibinfo {author} {\bibfnamefont {G.}~\bibnamefont {Sakkas}},\ }\href {\doibase 10.1103/PhysRevLett.129.201603} {\bibfield  {journal} {\bibinfo  {journal} {Phys. Rev. Lett.}\ }\textbf {\bibinfo {volume} {129}},\ \bibinfo {pages} {201603} (\bibinfo {year} {2022})},\ \Eprint {http://arxiv.org/abs/2203.17157} {arXiv:2203.17157} \BibitemShut {NoStop}%
\bibitem [{\citenamefont {Gabai}\ \emph {et~al.}(2024)\citenamefont {Gabai}, \citenamefont {Sever},\ and\ \citenamefont {Zhong}}]{Gabai:2023lax}%
  \BibitemOpen
  \bibfield  {author} {\bibinfo {author} {\bibfnamefont {B.}~\bibnamefont {Gabai}}, \bibinfo {author} {\bibfnamefont {A.}~\bibnamefont {Sever}}, \ and\ \bibinfo {author} {\bibfnamefont {D.-l.}\ \bibnamefont {Zhong}},\ }\href {\doibase 10.1007/JHEP03(2024)055} {\bibfield  {journal} {\bibinfo  {journal} {JHEP}\ }\textbf {\bibinfo {volume} {03}},\ \bibinfo {pages} {055} (\bibinfo {year} {2024})},\ \bibinfo {note} {[Erratum: JHEP 12, 083 (2024)]},\ \Eprint {http://arxiv.org/abs/2312.17132} {arXiv:2312.17132} \BibitemShut {NoStop}%
\bibitem [{Note2()}]{Note2}%
  \BibitemOpen
  \bibinfo {note} {We do not include here the possibly of breaking the transverse rotation group~\cite {Gukov:2006jk, Drukker:2008jm, Kravchuk:2025evf, vortex}, though do discuss it in \cite {whale}.}\BibitemShut {Stop}%
\bibitem [{\citenamefont {Drukker}\ \emph {et~al.}(2020{\natexlab{a}})\citenamefont {Drukker}, \citenamefont {Probst},\ and\ \citenamefont {Tr{\'e}panier}}]{Drukker:2020dcz}%
  \BibitemOpen
  \bibfield  {author} {\bibinfo {author} {\bibfnamefont {N.}~\bibnamefont {Drukker}}, \bibinfo {author} {\bibfnamefont {M.}~\bibnamefont {Probst}}, \ and\ \bibinfo {author} {\bibfnamefont {M.}~\bibnamefont {Tr{\'e}panier}},\ }\href {\doibase 10.1088/1751-8121/aba1b7} {\bibfield  {journal} {\bibinfo  {journal} {J. Phys. A}\ }\textbf {\bibinfo {volume} {53}},\ \bibinfo {pages} {365401} (\bibinfo {year} {2020}{\natexlab{a}})},\ \Eprint {http://arxiv.org/abs/2003.12372} {arXiv:2003.12372} \BibitemShut {NoStop}%
\bibitem [{\citenamefont {Drukker}\ \emph {et~al.}(2021)\citenamefont {Drukker}, \citenamefont {Probst},\ and\ \citenamefont {Tr{\'e}panier}}]{Drukker:2020atp}%
  \BibitemOpen
  \bibfield  {author} {\bibinfo {author} {\bibfnamefont {N.}~\bibnamefont {Drukker}}, \bibinfo {author} {\bibfnamefont {M.}~\bibnamefont {Probst}}, \ and\ \bibinfo {author} {\bibfnamefont {M.}~\bibnamefont {Tr{\'e}panier}},\ }\href {\doibase 10.1007/JHEP03(2021)261} {\bibfield  {journal} {\bibinfo  {journal} {JHEP}\ }\textbf {\bibinfo {volume} {03}},\ \bibinfo {pages} {261} (\bibinfo {year} {2021})},\ \Eprint {http://arxiv.org/abs/2009.10732} {arXiv:2009.10732} \BibitemShut {NoStop}%
\bibitem [{\citenamefont {Kutasov}(1989)}]{Kutasov:1988xb}%
  \BibitemOpen
  \bibfield  {author} {\bibinfo {author} {\bibfnamefont {D.}~\bibnamefont {Kutasov}},\ }\href {\doibase 10.1016/0370-2693(89)90028-2} {\bibfield  {journal} {\bibinfo  {journal} {Phys. Lett. B}\ }\textbf {\bibinfo {volume} {220}},\ \bibinfo {pages} {153} (\bibinfo {year} {1989})}\BibitemShut {NoStop}%
\bibitem [{\citenamefont {Friedan}\ and\ \citenamefont {Konechny}(2012)}]{Friedan:2012hi}%
  \BibitemOpen
  \bibfield  {author} {\bibinfo {author} {\bibfnamefont {D.}~\bibnamefont {Friedan}}\ and\ \bibinfo {author} {\bibfnamefont {A.}~\bibnamefont {Konechny}},\ }\href {\doibase 10.1007/JHEP09(2012)113} {\bibfield  {journal} {\bibinfo  {journal} {JHEP}\ }\textbf {\bibinfo {volume} {09}},\ \bibinfo {pages} {113} (\bibinfo {year} {2012})},\ \Eprint {http://arxiv.org/abs/1206.1749} {arXiv:1206.1749} \BibitemShut {NoStop}%
\bibitem [{\citenamefont {Bliard}\ \emph {et~al.}(2025)\citenamefont {Bliard}, \citenamefont {Correa}, \citenamefont {Lagares},\ and\ \citenamefont {Salazar~Landea}}]{Bliard:2024bcz}%
  \BibitemOpen
  \bibfield  {author} {\bibinfo {author} {\bibfnamefont {G.}~\bibnamefont {Bliard}}, \bibinfo {author} {\bibfnamefont {D.~H.}\ \bibnamefont {Correa}}, \bibinfo {author} {\bibfnamefont {M.}~\bibnamefont {Lagares}}, \ and\ \bibinfo {author} {\bibfnamefont {I.}~\bibnamefont {Salazar~Landea}},\ }\href {\doibase 10.1007/JHEP01(2025)131} {\bibfield  {journal} {\bibinfo  {journal} {JHEP}\ }\textbf {\bibinfo {volume} {01}},\ \bibinfo {pages} {131} (\bibinfo {year} {2025})},\ \Eprint {http://arxiv.org/abs/2410.02685} {arXiv:2410.02685} \BibitemShut {NoStop}%
\bibitem [{\citenamefont {Gimenez-Grau}\ and\ \citenamefont {Liendo}(2020)}]{Gimenez-Grau:2019hez}%
  \BibitemOpen
  \bibfield  {author} {\bibinfo {author} {\bibfnamefont {A.}~\bibnamefont {Gimenez-Grau}}\ and\ \bibinfo {author} {\bibfnamefont {P.}~\bibnamefont {Liendo}},\ }\href {\doibase 10.1007/JHEP03(2020)121} {\bibfield  {journal} {\bibinfo  {journal} {JHEP}\ }\textbf {\bibinfo {volume} {03}},\ \bibinfo {pages} {121} (\bibinfo {year} {2020})},\ \Eprint {http://arxiv.org/abs/1907.04345} {arXiv:1907.04345} \BibitemShut {NoStop}%
\bibitem [{\citenamefont {Pozzi}\ and\ \citenamefont {Trancanelli}(2024)}]{Pozzi:2024xnu}%
  \BibitemOpen
  \bibfield  {author} {\bibinfo {author} {\bibfnamefont {R.~G.}\ \bibnamefont {Pozzi}}\ and\ \bibinfo {author} {\bibfnamefont {D.}~\bibnamefont {Trancanelli}},\ }\href {\doibase 10.1103/PhysRevD.110.066006} {\bibfield  {journal} {\bibinfo  {journal} {Phys. Rev. D}\ }\textbf {\bibinfo {volume} {110}},\ \bibinfo {pages} {066006} (\bibinfo {year} {2024})},\ \Eprint {http://arxiv.org/abs/2406.13571} {arXiv:2406.13571} \BibitemShut {NoStop}%
\bibitem [{\citenamefont {Cavagli\`a}\ \emph {et~al.}(2022)\citenamefont {Cavagli\`a}, \citenamefont {Gromov}, \citenamefont {Julius},\ and\ \citenamefont {Preti}}]{Cavaglia:2022qpg}%
  \BibitemOpen
  \bibfield  {author} {\bibinfo {author} {\bibfnamefont {A.}~\bibnamefont {Cavagli\`a}}, \bibinfo {author} {\bibfnamefont {N.}~\bibnamefont {Gromov}}, \bibinfo {author} {\bibfnamefont {J.}~\bibnamefont {Julius}}, \ and\ \bibinfo {author} {\bibfnamefont {M.}~\bibnamefont {Preti}},\ }\href {\doibase 10.1007/JHEP05(2022)164} {\bibfield  {journal} {\bibinfo  {journal} {JHEP}\ }\textbf {\bibinfo {volume} {05}},\ \bibinfo {pages} {164} (\bibinfo {year} {2022})},\ \Eprint {http://arxiv.org/abs/2203.09556} {arXiv:2203.09556} \BibitemShut {NoStop}%
\bibitem [{\citenamefont {Sakkas}(2024)}]{Sakkas:2024dvm}%
  \BibitemOpen
  \bibfield  {author} {\bibinfo {author} {\bibfnamefont {G.}~\bibnamefont {Sakkas}},\ }\href@noop {} {\  (\bibinfo {year} {2024})},\ \Eprint {http://arxiv.org/abs/2403.05243} {arXiv:2403.05243} \BibitemShut {NoStop}%
\bibitem [{\citenamefont {Wilson}\ and\ \citenamefont {Fisher}(1972)}]{Wilson:1971dc}%
  \BibitemOpen
  \bibfield  {author} {\bibinfo {author} {\bibfnamefont {K.~G.}\ \bibnamefont {Wilson}}\ and\ \bibinfo {author} {\bibfnamefont {M.~E.}\ \bibnamefont {Fisher}},\ }\href {\doibase 10.1103/PhysRevLett.28.240} {\bibfield  {journal} {\bibinfo  {journal} {Phys. Rev. Lett.}\ }\textbf {\bibinfo {volume} {28}},\ \bibinfo {pages} {240} (\bibinfo {year} {1972})}\BibitemShut {NoStop}%
\bibitem [{\citenamefont {Allais}\ and\ \citenamefont {Sachdev}(2014)}]{Allais:2014fqa}%
  \BibitemOpen
  \bibfield  {author} {\bibinfo {author} {\bibfnamefont {A.}~\bibnamefont {Allais}}\ and\ \bibinfo {author} {\bibfnamefont {S.}~\bibnamefont {Sachdev}},\ }\href {\doibase 10.1103/PhysRevB.90.035131} {\bibfield  {journal} {\bibinfo  {journal} {Phys. Rev. B}\ }\textbf {\bibinfo {volume} {90}},\ \bibinfo {pages} {035131} (\bibinfo {year} {2014})},\ \Eprint {http://arxiv.org/abs/1406.3022} {arXiv:1406.3022} \BibitemShut {NoStop}%
\bibitem [{\citenamefont {Cuomo}\ \emph {et~al.}(2022)\citenamefont {Cuomo}, \citenamefont {Komargodski},\ and\ \citenamefont {Mezei}}]{Cuomo:2021kfm}%
  \BibitemOpen
  \bibfield  {author} {\bibinfo {author} {\bibfnamefont {G.}~\bibnamefont {Cuomo}}, \bibinfo {author} {\bibfnamefont {Z.}~\bibnamefont {Komargodski}}, \ and\ \bibinfo {author} {\bibfnamefont {M.}~\bibnamefont {Mezei}},\ }\href {\doibase 10.1007/JHEP02(2022)134} {\bibfield  {journal} {\bibinfo  {journal} {JHEP}\ }\textbf {\bibinfo {volume} {02}},\ \bibinfo {pages} {134} (\bibinfo {year} {2022})},\ \Eprint {http://arxiv.org/abs/2112.10634} {arXiv:2112.10634} \BibitemShut {NoStop}%
\bibitem [{\citenamefont {Gimenez-Grau}\ \emph {et~al.}(2022)\citenamefont {Gimenez-Grau}, \citenamefont {Lauria}, \citenamefont {Liendo},\ and\ \citenamefont {van Vliet}}]{Gimenez-Grau:2022czc}%
  \BibitemOpen
  \bibfield  {author} {\bibinfo {author} {\bibfnamefont {A.}~\bibnamefont {Gimenez-Grau}}, \bibinfo {author} {\bibfnamefont {E.}~\bibnamefont {Lauria}}, \bibinfo {author} {\bibfnamefont {P.}~\bibnamefont {Liendo}}, \ and\ \bibinfo {author} {\bibfnamefont {P.}~\bibnamefont {van Vliet}},\ }\href {\doibase 10.1007/JHEP11(2022)018} {\bibfield  {journal} {\bibinfo  {journal} {JHEP}\ }\textbf {\bibinfo {volume} {11}},\ \bibinfo {pages} {018} (\bibinfo {year} {2022})},\ \Eprint {http://arxiv.org/abs/2208.11715} {arXiv:2208.11715} \BibitemShut {NoStop}%
\bibitem [{\citenamefont {Artico}\ \emph {et~al.}(2025)\citenamefont {Artico}, \citenamefont {Barrat},\ and\ \citenamefont {Peveri}}]{Artico:2024wut}%
  \BibitemOpen
  \bibfield  {author} {\bibinfo {author} {\bibfnamefont {D.}~\bibnamefont {Artico}}, \bibinfo {author} {\bibfnamefont {J.}~\bibnamefont {Barrat}}, \ and\ \bibinfo {author} {\bibfnamefont {G.}~\bibnamefont {Peveri}},\ }\href {\doibase 10.1007/JHEP02(2025)190} {\bibfield  {journal} {\bibinfo  {journal} {JHEP}\ }\textbf {\bibinfo {volume} {02}},\ \bibinfo {pages} {190} (\bibinfo {year} {2025})},\ \Eprint {http://arxiv.org/abs/2410.08271} {arXiv:2410.08271} \BibitemShut {NoStop}%
\bibitem [{\citenamefont {Usyukina}\ and\ \citenamefont {Davydychev}(1993)}]{Usyukina:1992jd}%
  \BibitemOpen
  \bibfield  {author} {\bibinfo {author} {\bibfnamefont {N.~I.}\ \bibnamefont {Usyukina}}\ and\ \bibinfo {author} {\bibfnamefont {A.~I.}\ \bibnamefont {Davydychev}},\ }\href {\doibase 10.1016/0370-2693(93)91834-A} {\bibfield  {journal} {\bibinfo  {journal} {Phys. Lett. B}\ }\textbf {\bibinfo {volume} {298}},\ \bibinfo {pages} {363} (\bibinfo {year} {1993})}\BibitemShut {NoStop}%
\bibitem [{\citenamefont {Jensen}\ \emph {et~al.}(2018)\citenamefont {Jensen}, \citenamefont {Shaverin},\ and\ \citenamefont {Yarom}}]{Jensen:2017eof}%
  \BibitemOpen
  \bibfield  {author} {\bibinfo {author} {\bibfnamefont {K.}~\bibnamefont {Jensen}}, \bibinfo {author} {\bibfnamefont {E.}~\bibnamefont {Shaverin}}, \ and\ \bibinfo {author} {\bibfnamefont {A.}~\bibnamefont {Yarom}},\ }\href {\doibase 10.1007/JHEP01(2018)085} {\bibfield  {journal} {\bibinfo  {journal} {JHEP}\ }\textbf {\bibinfo {volume} {01}},\ \bibinfo {pages} {085} (\bibinfo {year} {2018})},\ \Eprint {http://arxiv.org/abs/1710.07299} {arXiv:1710.07299} \BibitemShut {NoStop}%
\bibitem [{\citenamefont {Choi}\ \emph {et~al.}(2025)\citenamefont {Choi}, \citenamefont {Ha}, \citenamefont {Kim}, \citenamefont {Kusuki}, \citenamefont {Ohyama},\ and\ \citenamefont {Ryu}}]{Choi:2025ebk}%
  \BibitemOpen
  \bibfield  {author} {\bibinfo {author} {\bibfnamefont {Y.}~\bibnamefont {Choi}}, \bibinfo {author} {\bibfnamefont {H.}~\bibnamefont {Ha}}, \bibinfo {author} {\bibfnamefont {D.}~\bibnamefont {Kim}}, \bibinfo {author} {\bibfnamefont {Y.}~\bibnamefont {Kusuki}}, \bibinfo {author} {\bibfnamefont {S.}~\bibnamefont {Ohyama}}, \ and\ \bibinfo {author} {\bibfnamefont {S.}~\bibnamefont {Ryu}},\ }\href@noop {} {\  (\bibinfo {year} {2025})},\ \Eprint {http://arxiv.org/abs/2507.12525} {arXiv:2507.12525} \BibitemShut {NoStop}%
\bibitem [{\citenamefont {Wen}(2025)}]{Wen:2025xka}%
  \BibitemOpen
  \bibfield  {author} {\bibinfo {author} {\bibfnamefont {X.}~\bibnamefont {Wen}},\ }\href@noop {} {\  (\bibinfo {year} {2025})},\ \Eprint {http://arxiv.org/abs/2507.12546} {arXiv:2507.12546} \BibitemShut {NoStop}%
\bibitem [{\citenamefont {Copetti}(2025)}]{Copetti:2025sym}%
  \BibitemOpen
  \bibfield  {author} {\bibinfo {author} {\bibfnamefont {C.}~\bibnamefont {Copetti}},\ }\href@noop {} {\  (\bibinfo {year} {2025})},\ \Eprint {http://arxiv.org/abs/2507.15466} {arXiv:2507.15466} \BibitemShut {NoStop}%
\bibitem [{\citenamefont {Drukker}\ \emph {et~al.}(2020{\natexlab{b}})\citenamefont {Drukker}, \citenamefont {Giombi}, \citenamefont {Tseytlin},\ and\ \citenamefont {Zhou}}]{Drukker:2020swu}%
  \BibitemOpen
  \bibfield  {author} {\bibinfo {author} {\bibfnamefont {N.}~\bibnamefont {Drukker}}, \bibinfo {author} {\bibfnamefont {S.}~\bibnamefont {Giombi}}, \bibinfo {author} {\bibfnamefont {A.~A.}\ \bibnamefont {Tseytlin}}, \ and\ \bibinfo {author} {\bibfnamefont {X.}~\bibnamefont {Zhou}},\ }\href {\doibase 10.1007/JHEP07(2020)101} {\bibfield  {journal} {\bibinfo  {journal} {JHEP}\ }\textbf {\bibinfo {volume} {07}},\ \bibinfo {pages} {101} (\bibinfo {year} {2020}{\natexlab{b}})},\ \Eprint {http://arxiv.org/abs/2004.04562} {arXiv:2004.04562} \BibitemShut {NoStop}%
\bibitem [{\citenamefont {Giombi}\ \emph {et~al.}(2017)\citenamefont {Giombi}, \citenamefont {Roiban},\ and\ \citenamefont {Tseytlin}}]{Giombi:2017cqn}%
  \BibitemOpen
  \bibfield  {author} {\bibinfo {author} {\bibfnamefont {S.}~\bibnamefont {Giombi}}, \bibinfo {author} {\bibfnamefont {R.}~\bibnamefont {Roiban}}, \ and\ \bibinfo {author} {\bibfnamefont {A.~A.}\ \bibnamefont {Tseytlin}},\ }\href {\doibase 10.1016/j.nuclphysb.2017.07.004} {\bibfield  {journal} {\bibinfo  {journal} {Nucl. Phys. B}\ }\textbf {\bibinfo {volume} {922}},\ \bibinfo {pages} {499} (\bibinfo {year} {2017})},\ \Eprint {http://arxiv.org/abs/1706.00756} {arXiv:1706.00756} \BibitemShut {NoStop}%
\bibitem [{\citenamefont {Giombi}\ and\ \citenamefont {Komatsu}(2018)}]{Giombi:2018qox}%
  \BibitemOpen
  \bibfield  {author} {\bibinfo {author} {\bibfnamefont {S.}~\bibnamefont {Giombi}}\ and\ \bibinfo {author} {\bibfnamefont {S.}~\bibnamefont {Komatsu}},\ }\href {\doibase 10.1007/JHEP05(2018)109} {\bibfield  {journal} {\bibinfo  {journal} {JHEP}\ }\textbf {\bibinfo {volume} {05}},\ \bibinfo {pages} {109} (\bibinfo {year} {2018})},\ \bibinfo {note} {[Erratum: JHEP 11, 123 (2018)]},\ \Eprint {http://arxiv.org/abs/1802.05201} {arXiv:1802.05201} \BibitemShut {NoStop}%
\bibitem [{\citenamefont {Liendo}\ \emph {et~al.}(2018)\citenamefont {Liendo}, \citenamefont {Meneghelli},\ and\ \citenamefont {Mitev}}]{Liendo:2018ukf}%
  \BibitemOpen
  \bibfield  {author} {\bibinfo {author} {\bibfnamefont {P.}~\bibnamefont {Liendo}}, \bibinfo {author} {\bibfnamefont {C.}~\bibnamefont {Meneghelli}}, \ and\ \bibinfo {author} {\bibfnamefont {V.}~\bibnamefont {Mitev}},\ }\href {\doibase 10.1007/JHEP10(2018)077} {\bibfield  {journal} {\bibinfo  {journal} {JHEP}\ }\textbf {\bibinfo {volume} {10}},\ \bibinfo {pages} {077} (\bibinfo {year} {2018})},\ \Eprint {http://arxiv.org/abs/1806.01862} {arXiv:1806.01862} \BibitemShut {NoStop}%
\bibitem [{\citenamefont {Kiryu}\ and\ \citenamefont {Komatsu}(2019)}]{Kiryu:2018phb}%
  \BibitemOpen
  \bibfield  {author} {\bibinfo {author} {\bibfnamefont {N.}~\bibnamefont {Kiryu}}\ and\ \bibinfo {author} {\bibfnamefont {S.}~\bibnamefont {Komatsu}},\ }\href {\doibase 10.1007/JHEP02(2019)090} {\bibfield  {journal} {\bibinfo  {journal} {JHEP}\ }\textbf {\bibinfo {volume} {02}},\ \bibinfo {pages} {090} (\bibinfo {year} {2019})},\ \Eprint {http://arxiv.org/abs/1812.04593} {arXiv:1812.04593} \BibitemShut {NoStop}%
\bibitem [{\citenamefont {Beccaria}\ \emph {et~al.}(2019)\citenamefont {Beccaria}, \citenamefont {Giombi},\ and\ \citenamefont {Tseytlin}}]{Beccaria:2019dws}%
  \BibitemOpen
  \bibfield  {author} {\bibinfo {author} {\bibfnamefont {M.}~\bibnamefont {Beccaria}}, \bibinfo {author} {\bibfnamefont {S.}~\bibnamefont {Giombi}}, \ and\ \bibinfo {author} {\bibfnamefont {A.~A.}\ \bibnamefont {Tseytlin}},\ }\href {\doibase 10.1007/JHEP05(2019)122} {\bibfield  {journal} {\bibinfo  {journal} {JHEP}\ }\textbf {\bibinfo {volume} {05}},\ \bibinfo {pages} {122} (\bibinfo {year} {2019})},\ \Eprint {http://arxiv.org/abs/1903.04365} {arXiv:1903.04365} \BibitemShut {NoStop}%
\bibitem [{\citenamefont {Bianchi}\ \emph {et~al.}(2020)\citenamefont {Bianchi}, \citenamefont {Bliard}, \citenamefont {Forini}, \citenamefont {Griguolo},\ and\ \citenamefont {Seminara}}]{Bianchi:2020hsz}%
  \BibitemOpen
  \bibfield  {author} {\bibinfo {author} {\bibfnamefont {L.}~\bibnamefont {Bianchi}}, \bibinfo {author} {\bibfnamefont {G.}~\bibnamefont {Bliard}}, \bibinfo {author} {\bibfnamefont {V.}~\bibnamefont {Forini}}, \bibinfo {author} {\bibfnamefont {L.}~\bibnamefont {Griguolo}}, \ and\ \bibinfo {author} {\bibfnamefont {D.}~\bibnamefont {Seminara}},\ }\href {\doibase 10.1007/JHEP08(2020)143} {\bibfield  {journal} {\bibinfo  {journal} {JHEP}\ }\textbf {\bibinfo {volume} {08}},\ \bibinfo {pages} {143} (\bibinfo {year} {2020})},\ \Eprint {http://arxiv.org/abs/2004.07849} {arXiv:2004.07849} \BibitemShut {NoStop}%
\bibitem [{\citenamefont {Ferrero}\ and\ \citenamefont {Meneghelli}(2021)}]{Ferrero:2021bsb}%
  \BibitemOpen
  \bibfield  {author} {\bibinfo {author} {\bibfnamefont {P.}~\bibnamefont {Ferrero}}\ and\ \bibinfo {author} {\bibfnamefont {C.}~\bibnamefont {Meneghelli}},\ }\href {\doibase 10.1103/PhysRevD.104.L081703} {\bibfield  {journal} {\bibinfo  {journal} {Phys. Rev. D}\ }\textbf {\bibinfo {volume} {104}},\ \bibinfo {pages} {L081703} (\bibinfo {year} {2021})},\ \Eprint {http://arxiv.org/abs/2103.10440} {arXiv:2103.10440} \BibitemShut {NoStop}%
\bibitem [{\citenamefont {Giombi}\ \emph {et~al.}(2022{\natexlab{a}})\citenamefont {Giombi}, \citenamefont {Komatsu},\ and\ \citenamefont {Offertaler}}]{Giombi:2021zfb}%
  \BibitemOpen
  \bibfield  {author} {\bibinfo {author} {\bibfnamefont {S.}~\bibnamefont {Giombi}}, \bibinfo {author} {\bibfnamefont {S.}~\bibnamefont {Komatsu}}, \ and\ \bibinfo {author} {\bibfnamefont {B.}~\bibnamefont {Offertaler}},\ }\href {\doibase 10.1007/JHEP03(2022)020} {\bibfield  {journal} {\bibinfo  {journal} {JHEP}\ }\textbf {\bibinfo {volume} {03}},\ \bibinfo {pages} {020} (\bibinfo {year} {2022}{\natexlab{a}})},\ \Eprint {http://arxiv.org/abs/2110.13126} {arXiv:2110.13126} \BibitemShut {NoStop}%
\bibitem [{\citenamefont {Giombi}\ \emph {et~al.}(2022{\natexlab{b}})\citenamefont {Giombi}, \citenamefont {Komatsu},\ and\ \citenamefont {Offertaler}}]{Giombi:2022anm}%
  \BibitemOpen
  \bibfield  {author} {\bibinfo {author} {\bibfnamefont {S.}~\bibnamefont {Giombi}}, \bibinfo {author} {\bibfnamefont {S.}~\bibnamefont {Komatsu}}, \ and\ \bibinfo {author} {\bibfnamefont {B.}~\bibnamefont {Offertaler}},\ }\href {\doibase 10.1007/JHEP08(2022)011} {\bibfield  {journal} {\bibinfo  {journal} {JHEP}\ }\textbf {\bibinfo {volume} {08}},\ \bibinfo {pages} {011} (\bibinfo {year} {2022}{\natexlab{b}})},\ \Eprint {http://arxiv.org/abs/2202.07627} {arXiv:2202.07627} \BibitemShut {NoStop}%
\bibitem [{\citenamefont {Bliard}(2023)}]{Bliard:2023zpe}%
  \BibitemOpen
  \bibfield  {author} {\bibinfo {author} {\bibfnamefont {G.~J.~S.}\ \bibnamefont {Bliard}},\ }\emph {\bibinfo {title} {{Perturbative and non-perturbative analysis of defect correlators in AdS/CFT}}},\ \href {\doibase 10.18452/27559} {Ph.D. thesis},\ \bibinfo  {school} {Humboldt U., Berlin} (\bibinfo {year} {2023}),\ \Eprint {http://arxiv.org/abs/2310.18137} {arXiv:2310.18137} \BibitemShut {NoStop}%
\bibitem [{\citenamefont {Ferrero}\ and\ \citenamefont {Meneghelli}(2024{\natexlab{a}})}]{Ferrero:2023znz}%
  \BibitemOpen
  \bibfield  {author} {\bibinfo {author} {\bibfnamefont {P.}~\bibnamefont {Ferrero}}\ and\ \bibinfo {author} {\bibfnamefont {C.}~\bibnamefont {Meneghelli}},\ }\href {\doibase 10.1007/JHEP05(2024)090} {\bibfield  {journal} {\bibinfo  {journal} {JHEP}\ }\textbf {\bibinfo {volume} {05}},\ \bibinfo {pages} {090} (\bibinfo {year} {2024}{\natexlab{a}})},\ \Eprint {http://arxiv.org/abs/2312.12550} {arXiv:2312.12550} \BibitemShut {NoStop}%
\bibitem [{\citenamefont {Ferrero}\ and\ \citenamefont {Meneghelli}(2024{\natexlab{b}})}]{Ferrero:2023gnu}%
  \BibitemOpen
  \bibfield  {author} {\bibinfo {author} {\bibfnamefont {P.}~\bibnamefont {Ferrero}}\ and\ \bibinfo {author} {\bibfnamefont {C.}~\bibnamefont {Meneghelli}},\ }\href {\doibase 10.1007/JHEP06(2024)010} {\bibfield  {journal} {\bibinfo  {journal} {JHEP}\ }\textbf {\bibinfo {volume} {06}},\ \bibinfo {pages} {010} (\bibinfo {year} {2024}{\natexlab{b}})},\ \Eprint {http://arxiv.org/abs/2312.12551} {arXiv:2312.12551} \BibitemShut {NoStop}%
\bibitem [{Note3()}]{Note3}%
  \BibitemOpen
  \bibinfo {note} {As already noted in \cite {Gabai:2025zcs, Girault:2025kzt}.}\BibitemShut {Stop}%
\bibitem [{\citenamefont {Alday}\ and\ \citenamefont {Maldacena}(2007)}]{Alday:2007he}%
  \BibitemOpen
  \bibfield  {author} {\bibinfo {author} {\bibfnamefont {L.~F.}\ \bibnamefont {Alday}}\ and\ \bibinfo {author} {\bibfnamefont {J.}~\bibnamefont {Maldacena}},\ }\href {\doibase 10.1088/1126-6708/2007/11/068} {\bibfield  {journal} {\bibinfo  {journal} {JHEP}\ }\textbf {\bibinfo {volume} {11}},\ \bibinfo {pages} {068} (\bibinfo {year} {2007})},\ \Eprint {http://arxiv.org/abs/0710.1060} {arXiv:0710.1060} \BibitemShut {NoStop}%
\bibitem [{\citenamefont {Polchinski}\ and\ \citenamefont {Sully}(2011)}]{Polchinski:2011im}%
  \BibitemOpen
  \bibfield  {author} {\bibinfo {author} {\bibfnamefont {J.}~\bibnamefont {Polchinski}}\ and\ \bibinfo {author} {\bibfnamefont {J.}~\bibnamefont {Sully}},\ }\href {\doibase 10.1007/JHEP10(2011)059} {\bibfield  {journal} {\bibinfo  {journal} {JHEP}\ }\textbf {\bibinfo {volume} {10}},\ \bibinfo {pages} {059} (\bibinfo {year} {2011})},\ \Eprint {http://arxiv.org/abs/1104.5077} {arXiv:1104.5077} \BibitemShut {NoStop}%
\bibitem [{\citenamefont {Deser}\ and\ \citenamefont {Schwimmer}(1993)}]{Deser:1993yx}%
  \BibitemOpen
  \bibfield  {author} {\bibinfo {author} {\bibfnamefont {S.}~\bibnamefont {Deser}}\ and\ \bibinfo {author} {\bibfnamefont {A.}~\bibnamefont {Schwimmer}},\ }\href {\doibase 10.1016/0370-2693(93)90934-A} {\bibfield  {journal} {\bibinfo  {journal} {Phys. Lett. B}\ }\textbf {\bibinfo {volume} {309}},\ \bibinfo {pages} {279} (\bibinfo {year} {1993})},\ \Eprint {http://arxiv.org/abs/hep-th/9302047} {hep-th/9302047} \BibitemShut {NoStop}%
\bibitem [{\citenamefont {Graham}\ and\ \citenamefont {Witten}(1999)}]{Graham:1999pm}%
  \BibitemOpen
  \bibfield  {author} {\bibinfo {author} {\bibfnamefont {C.~R.}\ \bibnamefont {Graham}}\ and\ \bibinfo {author} {\bibfnamefont {E.}~\bibnamefont {Witten}},\ }\href {\doibase 10.1016/S0550-3213(99)00055-3} {\bibfield  {journal} {\bibinfo  {journal} {Nucl. Phys. B}\ }\textbf {\bibinfo {volume} {546}},\ \bibinfo {pages} {52} (\bibinfo {year} {1999})},\ \Eprint {http://arxiv.org/abs/hep-th/9901021} {arXiv:hep-th/9901021} \BibitemShut {NoStop}%
\bibitem [{\citenamefont {Chalabi}\ \emph {et~al.}(2022)\citenamefont {Chalabi}, \citenamefont {Herzog}, \citenamefont {O'Bannon}, \citenamefont {Robinson},\ and\ \citenamefont {Sisti}}]{Chalabi:2021jud}%
  \BibitemOpen
  \bibfield  {author} {\bibinfo {author} {\bibfnamefont {A.}~\bibnamefont {Chalabi}}, \bibinfo {author} {\bibfnamefont {C.~P.}\ \bibnamefont {Herzog}}, \bibinfo {author} {\bibfnamefont {A.}~\bibnamefont {O'Bannon}}, \bibinfo {author} {\bibfnamefont {B.}~\bibnamefont {Robinson}}, \ and\ \bibinfo {author} {\bibfnamefont {J.}~\bibnamefont {Sisti}},\ }\href {\doibase 10.1007/JHEP02(2022)166} {\bibfield  {journal} {\bibinfo  {journal} {JHEP}\ }\textbf {\bibinfo {volume} {02}},\ \bibinfo {pages} {166} (\bibinfo {year} {2022})},\ \Eprint {http://arxiv.org/abs/2111.14713} {arXiv:2111.14713} \BibitemShut {NoStop}%
\bibitem [{\citenamefont {Giombi}\ and\ \citenamefont {Tseytlin}(2023)}]{Giombi:2023vzu}%
  \BibitemOpen
  \bibfield  {author} {\bibinfo {author} {\bibfnamefont {S.}~\bibnamefont {Giombi}}\ and\ \bibinfo {author} {\bibfnamefont {A.~A.}\ \bibnamefont {Tseytlin}},\ }\href {\doibase 10.1103/PhysRevLett.130.201601} {\bibfield  {journal} {\bibinfo  {journal} {Phys. Rev. Lett.}\ }\textbf {\bibinfo {volume} {130}},\ \bibinfo {pages} {201601} (\bibinfo {year} {2023})},\ \Eprint {http://arxiv.org/abs/2303.15207} {arXiv:2303.15207} \BibitemShut {NoStop}%
\bibitem [{\citenamefont {Rodgers}(2019)}]{Rodgers:2018mvq}%
  \BibitemOpen
  \bibfield  {author} {\bibinfo {author} {\bibfnamefont {R.}~\bibnamefont {Rodgers}},\ }\href {\doibase 10.1007/JHEP03(2019)092} {\bibfield  {journal} {\bibinfo  {journal} {JHEP}\ }\textbf {\bibinfo {volume} {03}},\ \bibinfo {pages} {092} (\bibinfo {year} {2019})},\ \Eprint {http://arxiv.org/abs/1811.12375} {arXiv:1811.12375} \BibitemShut {NoStop}%
\bibitem [{\citenamefont {Estes}\ \emph {et~al.}(2019)\citenamefont {Estes}, \citenamefont {Krym}, \citenamefont {O'Bannon}, \citenamefont {Robinson},\ and\ \citenamefont {Rodgers}}]{Estes:2018tnu}%
  \BibitemOpen
  \bibfield  {author} {\bibinfo {author} {\bibfnamefont {J.}~\bibnamefont {Estes}}, \bibinfo {author} {\bibfnamefont {D.}~\bibnamefont {Krym}}, \bibinfo {author} {\bibfnamefont {A.}~\bibnamefont {O'Bannon}}, \bibinfo {author} {\bibfnamefont {B.}~\bibnamefont {Robinson}}, \ and\ \bibinfo {author} {\bibfnamefont {R.}~\bibnamefont {Rodgers}},\ }\href {\doibase 10.1007/JHEP05(2019)032} {\bibfield  {journal} {\bibinfo  {journal} {JHEP}\ }\textbf {\bibinfo {volume} {05}},\ \bibinfo {pages} {032} (\bibinfo {year} {2019})},\ \Eprint {http://arxiv.org/abs/1812.00923} {arXiv:1812.00923} \BibitemShut {NoStop}%
\bibitem [{\citenamefont {Jensen}\ \emph {et~al.}(2019)\citenamefont {Jensen}, \citenamefont {O'Bannon}, \citenamefont {Robinson},\ and\ \citenamefont {Rodgers}}]{Jensen:2018rxu}%
  \BibitemOpen
  \bibfield  {author} {\bibinfo {author} {\bibfnamefont {K.}~\bibnamefont {Jensen}}, \bibinfo {author} {\bibfnamefont {A.}~\bibnamefont {O'Bannon}}, \bibinfo {author} {\bibfnamefont {B.}~\bibnamefont {Robinson}}, \ and\ \bibinfo {author} {\bibfnamefont {R.}~\bibnamefont {Rodgers}},\ }\href {\doibase 10.1103/PhysRevLett.122.241602} {\bibfield  {journal} {\bibinfo  {journal} {Phys. Rev. Lett.}\ }\textbf {\bibinfo {volume} {122}},\ \bibinfo {pages} {241602} (\bibinfo {year} {2019})},\ \Eprint {http://arxiv.org/abs/1812.08745} {arXiv:1812.08745} \BibitemShut {NoStop}%
\bibitem [{\citenamefont {Chalabi}\ \emph {et~al.}(2020)\citenamefont {Chalabi}, \citenamefont {O'Bannon}, \citenamefont {Robinson},\ and\ \citenamefont {Sisti}}]{Chalabi:2020iie}%
  \BibitemOpen
  \bibfield  {author} {\bibinfo {author} {\bibfnamefont {A.}~\bibnamefont {Chalabi}}, \bibinfo {author} {\bibfnamefont {A.}~\bibnamefont {O'Bannon}}, \bibinfo {author} {\bibfnamefont {B.}~\bibnamefont {Robinson}}, \ and\ \bibinfo {author} {\bibfnamefont {J.}~\bibnamefont {Sisti}},\ }\href {\doibase 10.1007/JHEP05(2020)095} {\bibfield  {journal} {\bibinfo  {journal} {JHEP}\ }\textbf {\bibinfo {volume} {05}},\ \bibinfo {pages} {095} (\bibinfo {year} {2020})},\ \Eprint {http://arxiv.org/abs/2003.02857} {arXiv:2003.02857} \BibitemShut {NoStop}%
\bibitem [{\citenamefont {Wang}(2021)}]{Wang:2020xkc}%
  \BibitemOpen
  \bibfield  {author} {\bibinfo {author} {\bibfnamefont {Y.}~\bibnamefont {Wang}},\ }\href {\doibase 10.1007/JHEP11(2021)122} {\bibfield  {journal} {\bibinfo  {journal} {JHEP}\ }\textbf {\bibinfo {volume} {11}},\ \bibinfo {pages} {122} (\bibinfo {year} {2021})},\ \Eprint {http://arxiv.org/abs/2012.06574} {arXiv:2012.06574} \BibitemShut {NoStop}%
\bibitem [{\citenamefont {Tr\'epanier}(2023)}]{Trepanier:2023tvb}%
  \BibitemOpen
  \bibfield  {author} {\bibinfo {author} {\bibfnamefont {M.}~\bibnamefont {Tr\'epanier}},\ }\href {\doibase 10.1007/JHEP09(2023)074} {\bibfield  {journal} {\bibinfo  {journal} {JHEP}\ }\textbf {\bibinfo {volume} {09}},\ \bibinfo {pages} {074} (\bibinfo {year} {2023})},\ \Eprint {http://arxiv.org/abs/2305.10486} {arXiv:2305.10486} \BibitemShut {NoStop}%
\bibitem [{\citenamefont {Giombi}\ and\ \citenamefont {Liu}(2023)}]{Giombi:2023dqs}%
  \BibitemOpen
  \bibfield  {author} {\bibinfo {author} {\bibfnamefont {S.}~\bibnamefont {Giombi}}\ and\ \bibinfo {author} {\bibfnamefont {B.}~\bibnamefont {Liu}},\ }\href {\doibase 10.1007/JHEP12(2023)004} {\bibfield  {journal} {\bibinfo  {journal} {JHEP}\ }\textbf {\bibinfo {volume} {12}},\ \bibinfo {pages} {004} (\bibinfo {year} {2023})},\ \Eprint {http://arxiv.org/abs/2305.11402} {arXiv:2305.11402} \BibitemShut {NoStop}%
\bibitem [{\citenamefont {Raviv-Moshe}\ and\ \citenamefont {Zhong}(2023)}]{Raviv-Moshe:2023yvq}%
  \BibitemOpen
  \bibfield  {author} {\bibinfo {author} {\bibfnamefont {A.}~\bibnamefont {Raviv-Moshe}}\ and\ \bibinfo {author} {\bibfnamefont {S.}~\bibnamefont {Zhong}},\ }\href {\doibase 10.1007/JHEP08(2023)143} {\bibfield  {journal} {\bibinfo  {journal} {JHEP}\ }\textbf {\bibinfo {volume} {08}},\ \bibinfo {pages} {143} (\bibinfo {year} {2023})},\ \Eprint {http://arxiv.org/abs/2305.11370} {arXiv:2305.11370} \BibitemShut {NoStop}%
\bibitem [{\citenamefont {Diatlyk}\ \emph {et~al.}(2025)\citenamefont {Diatlyk}, \citenamefont {Sun},\ and\ \citenamefont {Wang}}]{Diatlyk:2024ngd}%
  \BibitemOpen
  \bibfield  {author} {\bibinfo {author} {\bibfnamefont {O.}~\bibnamefont {Diatlyk}}, \bibinfo {author} {\bibfnamefont {Z.}~\bibnamefont {Sun}}, \ and\ \bibinfo {author} {\bibfnamefont {Y.}~\bibnamefont {Wang}},\ }\href {\doibase 10.1007/JHEP06(2025)131} {\bibfield  {journal} {\bibinfo  {journal} {JHEP}\ }\textbf {\bibinfo {volume} {06}},\ \bibinfo {pages} {131} (\bibinfo {year} {2025})},\ \Eprint {http://arxiv.org/abs/2411.16522} {arXiv:2411.16522} \BibitemShut {NoStop}%
\bibitem [{\citenamefont {Gukov}\ and\ \citenamefont {Witten}(2006)}]{Gukov:2006jk}%
  \BibitemOpen
  \bibfield  {author} {\bibinfo {author} {\bibfnamefont {S.}~\bibnamefont {Gukov}}\ and\ \bibinfo {author} {\bibfnamefont {E.}~\bibnamefont {Witten}},\ }\href@noop {} {\  (\bibinfo {year} {2006})},\ \Eprint {http://arxiv.org/abs/hep-th/0612073} {arXiv:hep-th/0612073} \BibitemShut {NoStop}%
\bibitem [{\citenamefont {Drukker}\ \emph {et~al.}(2009)\citenamefont {Drukker}, \citenamefont {Gomis},\ and\ \citenamefont {Young}}]{Drukker:2008jm}%
  \BibitemOpen
  \bibfield  {author} {\bibinfo {author} {\bibfnamefont {N.}~\bibnamefont {Drukker}}, \bibinfo {author} {\bibfnamefont {J.}~\bibnamefont {Gomis}}, \ and\ \bibinfo {author} {\bibfnamefont {D.}~\bibnamefont {Young}},\ }\href {\doibase 10.1088/1126-6708/2009/03/004} {\bibfield  {journal} {\bibinfo  {journal} {JHEP}\ }\textbf {\bibinfo {volume} {03}},\ \bibinfo {pages} {004} (\bibinfo {year} {2009})},\ \Eprint {http://arxiv.org/abs/0810.4344} {arXiv:0810.4344} \BibitemShut {NoStop}%
\bibitem [{\citenamefont {Kravchuk}\ and\ \citenamefont {Radcliffe}(2025)}]{Kravchuk:2025evf}%
  \BibitemOpen
  \bibfield  {author} {\bibinfo {author} {\bibfnamefont {P.}~\bibnamefont {Kravchuk}}\ and\ \bibinfo {author} {\bibfnamefont {A.}~\bibnamefont {Radcliffe}},\ }\href@noop {} {\  (\bibinfo {year} {2025})},\ \Eprint {http://arxiv.org/abs/2510.02281} {arXiv:2510.02281} \BibitemShut {NoStop}%
\bibitem [{\citenamefont {Drukker}\ \emph {et~al.}(2025{\natexlab{b}})\citenamefont {Drukker}, \citenamefont {Kong}, \citenamefont {Shahpo},\ and\ \citenamefont {Tr\'epanier}}]{vortex}%
  \BibitemOpen
  \bibfield  {author} {\bibinfo {author} {\bibfnamefont {N.}~\bibnamefont {Drukker}}, \bibinfo {author} {\bibfnamefont {Z.}~\bibnamefont {Kong}}, \bibinfo {author} {\bibfnamefont {O.}~\bibnamefont {Shahpo}}, \ and\ \bibinfo {author} {\bibfnamefont {M.}~\bibnamefont {Tr\'epanier}},\ }\href@noop {} {\  (\bibinfo {year} {2025}{\natexlab{b}})},\ \bibinfo {note} {in progress}\BibitemShut {NoStop}%
\end{thebibliography}%
\end{document}